\documentclass[]{jfm}

\usepackage{graphicx}
\usepackage{newtxtext}
\usepackage{newtxmath}
\usepackage{natbib}
\usepackage{hyperref}
\hypersetup{
    colorlinks = true,
    linkcolor  = blue,
    urlcolor   = blue,
    citecolor  = blue,
}

\usepackage{epstopdf, epsfig}
\usepackage{amsmath}
\usepackage{psfrag}
\usepackage{subfigure}
\usepackage{mathrsfs}
\usepackage{color}
\usepackage{overpic}
\usepackage{comment}
\usepackage{float}
\usepackage{multirow}
\usepackage{amssymb}
\usepackage{makecell}

\usepackage{empheq}
\usepackage{nccmath}
\usepackage{tcolorbox}

\shorttitle{Integral relations for annular flows}
\shortauthor{P. Ricco}

\title{Integral relations for the skin-friction coefficients and other quantities of annular flows}

\author{Pierre Ricco\corresp{\email{p.ricco@sheffield.ac.uk}}}

\affiliation{School of Mechanical, Aerospace and Civil Engineering, The University of Sheffield, Sheffield S1 3JD, UK}

\begin{document}

\maketitle

\begin{abstract}
Integral identities relating the skin-friction coefficients and the Reynolds shear stresses in annular cylindrical flows are derived. In the pressure-driven case, an identity for the radial location of the maximum streamwise velocity is obtained and, in the cylinder-driven case, an identity for the bulk velocity is found. The formulas are used to analyse existing numerical and experimental data and simplify to classical channel-flow and pipe-flow identities in the limiting cases of vanishingly small and infinitely large radius ratios.
\end{abstract}

\vspace{-10mm}

\section{Introduction}

A major breakthrough in the analysis of turbulence was the decomposition of the flow into the mean and fluctuating quantities, which led to the mean-flow Navier--Stokes equations containing the averaged products of the velocity fluctuations \citep{reynolds-1895}. In fully developed canonical turbulent flows, the Reynolds shear stresses, involving the streamwise and wall-normal velocity fluctuations, distil the effect of turbulence on the flow dynamics. A central objective in turbulence research has been to quantify the impact of the Reynolds shear stresses on quantities of fluid engineering interest, such as the pressure gradient and the friction drag. \citet{fukagata-iwamoto-kasagi-2002} (FIK) succeeded in this regard as they discovered identities that, in their simplest form, express the turbulent skin-friction coefficient of classical flows in terms of a weighted wall-normal integral of the Reynolds shear stresses. An essential feature of these identities is that the laminar coefficient is isolated, i.e. the contribution of the Reynolds shear stresses to the skin-friction coefficient is quantified precisely. Three flows were considered by FIK: planar channel flows, circular pipe flows and free-stream boundary layers.

Our primary objective is to derive FIK-type integral identities for fully developed annular flows between concentric circular cylinders driven either by a streamwise pressure gradient or by the streamwise motion of the inner cylinder. Understanding and predicting such flows is of importance in fluid engineering applications, such as heat exchangers, cooling systems for nuclear reactors and machinery for petroleum extraction. Annular flows also represent a prototype of flows subjected simultaneously to concave and convex transverse curvatures. For pressure-driven annular flows, we also find an identity that relates the radial location of the maximum streamwise mean velocity to integrals involving the Reynolds shear stresses. For cylinder-driven annular flows, we report an identity for the bulk velocity. An important distinction from the channel and pipe flows studied by FIK is the difference between the skin-friction coefficients of the inner and outer cylinders. Experimental and numerical data are analysed and discussed using the new identities.
\section{Flow features and averaged quantities}
\begin{figure}
\centering
\includegraphics[width=6.7cm]{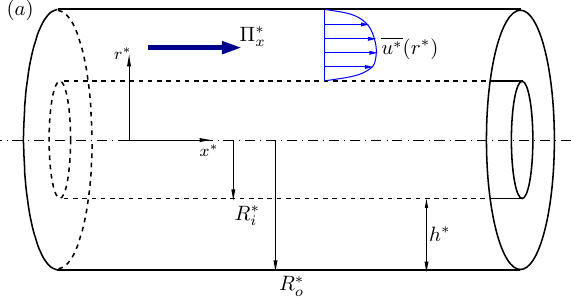}
\includegraphics[width=6.7cm]{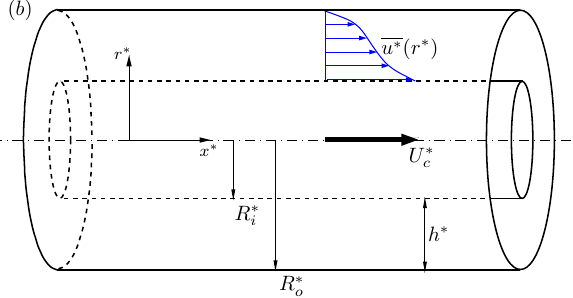}
\vspace{-5mm}
\caption{Schematics of the flow domains: (a) Pressure-driven annular flow, (b) Cylinder-driven annular flow.}
\label{fig:domain}
\end{figure}
Figure \ref{fig:domain} illustrates the two annular incompressible flows considered in this study. In the first configuration, the cylinders are stationary and the flow is driven by a streamwise pressure gradient $\Pi_x^*$. In the second configuration, the inner cylinder moves at a constant velocity $U_c^*$ along the streamwise direction, the outer cylinder is stationary and no pressure gradient exists. The superscript asterisk indicates dimensional quantities. Quantities without the superscript asterisk are non-dimensional. The annular gap is $h^*=R_o^*-R_i^*$, where $R_o^*$ and $R_i^*$ are the outer and inner radii. The operator $\overline{\\ \cdot \\}$ indicates averaging along the streamwise $x^*$ direction, the azimuthal $\theta$ direction and over time $t^*$. The flows are assumed to be statistically fully developed, i.e. the velocity statistics depend only on the radial direction $r^*$ and are independent of $\theta$, $x^*$ and $t^*$. Flow quantities are decomposed as $q^* = \overline{q^*} + {q'}^{*}$. The velocity components are $u^*,v^*,w^*$ along $x^*,r^*,\theta,$ respectively.
The bulk velocity is
\begin{equation}
\label{eq:bulk-dimensional}
U_b^* = \frac{2}{R_o^{*2}-R_i^{*2}} \int_{R_i^*}^{R_o^*} \overline{u^*}(r^*) r^* \mathrm{d}r^*.
\end{equation}

\subsection{Pressure-driven annular flow}

For the pressure-driven annular flow, the velocity scales as $\boldsymbol{u}=\{u,v,w\}=\boldsymbol{u}^*/U_b^*$ (where $U_b^*$ is a known quantity), the radial direction scales as $r = r^*/h^*$ and the pressure scales as $p = p^*/(\rho^* U_b^{*2})$, where $\rho^*$ is the density of the fluid. The Reynolds number is $Re_b = U_b^* h^*/\nu^*$, where $\nu^*$ is the kinematic viscosity of the fluid. Signed definitions of the skin-friction coefficients are used, i.e. the coefficients retain the sign of the velocity gradients.
The skin-friction coefficient for the mean wall-shear stress $\tau_{w,i}^*$ of the inner cylinder is
\begin{equation}
\label{eq:inner-cf-def}
C_{f,i} = \frac{2 \tau_{w,i}^*}{\rho^* U_b^{*2}}
= \frac{2 \nu^*}{U_b^{*2}} \left.\frac{\mathrm{d} \overline{u^*}}{\mathrm{d} r^*}\right|_{r^*=R_i^*}
= \frac{2}{Re_b} \left.\frac{\mathrm{d}\overline{u}}{\mathrm{d} r}\right|_{r=R_i} > 0.
\end{equation}
The skin-friction coefficient for the mean wall-shear stress $\tau_{w,o}^*$ of the {outer } cylinder is
\begin{equation}
\label{eq:outer-cf-def}
C_{f,o}
= \frac{2 \tau_{w,o}^*}{\rho^* U_b^{*2}}
= \frac{2 \nu^*}{U_b^{*2}} \left.\frac{\mathrm{d}\overline{u^*}}{\mathrm{d} r^*}\right|_{r^*=R_o^*}
= \frac{2}{Re_b} \left.\frac{\mathrm{d}\overline{u}}{\mathrm{d} r}\right|_{r=R_o} < 0.
\end{equation}
The average skin-friction coefficient is $C_f = 2 \mathcal{D}^*/\left(\mathcal{A}^* \rho^* U_b^{*2}\right)>0,$ where $\mathcal{D}^*$ is the drag exerted by the annular flow on the wetted area $\mathcal{A}^*$. In non-dimensional form, equation \eqref{eq:bulk-dimensional} is
\begin{equation}
\label{eq:bulk-pressure}
\int_{R_i}^{R_o} \overline{u}(r) r \mathrm{d}r = \frac{R_o^{2}-R_i^{2}}{2}.
\end{equation}

\subsection{Cylinder-driven annular flow}

For the cylinder-driven annular flow, the velocity scales as $\boldsymbol{u}=\{u,v,w\}=\boldsymbol{u}^*/U_c^*$ and the pressure scales as $p = p^*/(\rho^* U_c^{*2})$. The Reynolds number is $Re_c = U_c^* h^*/\nu^*$. The skin-friction coefficients for the cylinder-driven flow are defined analogously to those for the pressure-driven flow, except that the inner-cylinder velocity $U_c^*$ is used as the velocity scale instead of $U_b^*$. The skin-friction coefficients are termed $C_{f,i,c}$ and $C_{f,o,c}$. In non-dimensional form, equation \eqref{eq:bulk-dimensional} is
\begin{equation}
\label{eq:bulk-cylinder}
U_{b,c} = \dfrac{U_b^*}{U_c^*} = \frac{2}{R_o^{2}-R_i^{2}} \int_{R_i}^{R_o} \overline{u}(r) r \mathrm{d}r.
\end{equation}

\subsection{Nomenclature of pressure gradients and skin-friction coefficients}

Table \ref{tab:nomenclature} summarises the pressure gradients and the skin-friction coefficients for the flow configurations (pressure-driven flow and cylinder-driven flow) and regimes (laminar flows and transitional/turbulent flow).

\begin{table}
\centering
\begin{tabular}{c|cl|cl}
& \multicolumn{2}{c|}{Pressure-driven} & \multicolumn{2}{c}{Cylinder-driven} \\
\hline
\multirow{4}{*}{\begin{tabular}{@{}c@{}}Laminar\end{tabular}}
& $\Pi_{x,lam}$$<$0   & $x$-pressure gradient & $C_{f,i,c,lam}$$<$0 & Inner-cylinder SFC    \\
& $C_{f,lam}$$>$0     & Average SFC           & $C_{f,o,c,lam}$$<$0 & Outer-cylinder SFC    \\
& $C_{f,i,lam}$$>$0   & Inner-cylinder SFC    &                 &                           \\
& $C_{f,o,lam}$$<$0   & Outer-cylinder SFC    &                 &                           \\
\hline
\multirow{4}{*}{\begin{tabular}{@{}c@{}}Transitional/\\ Turbulent\end{tabular}}
& $\Pi_x$$<$0    & $x$-pressure gradient & $C_{f,c}$$<$0    & Average SFC                   \\
& $C_f$$>$0      & Average SFC           & $C_{f,i,c}$$<$0  & Inner-cylinder SFC            \\
& $C_{f,i}$$>$0  & Inner-cylinder SFC    & $C_{f,o,c}$$<$0  & Outer-cylinder SFC            \\
& $C_{f,o}$$<$0 & Outer-cylinder SFC    & $C_{f,c.uv}$$<$0 & Average SFC computed           \\
& $C_{f,*,uv}$$\lessgtr$0 & SFCs computed    &  & via \eqref{eq:cf-w}                       \\
&                         & via \eqref{eq:cf-turbulent}-\eqref{eq:cfo-turbulent} &  &    \\
&                         & (sign depends on whether &  &    \\
&                         & it is average, inner-cylinder or &  &    \\
&                         & outer-cylinder) &  &    \\
\hline
\end{tabular}
\caption{Nomenclature for the pressure gradients and skin-friction coefficients (SFC).}
\label{tab:nomenclature}
\end{table}

\section{Results for pressure-driven annular flow}

\subsection{Relations among the pressure gradient and the skin-friction coefficients}

By recalling that the velocity statistics depend on $r$ only and by integrating the averaged radial momentum equation, the averaged streamwise pressure gradient $\Pi_x<0$ is shown to be a constant. The averaged $x$-momentum equation becomes:
\begin{equation}
\label{eq:x-momentum}
    -\Pi_x = \frac{1}{r}\frac{\mathrm{d}}{\mathrm{d} r}\left(r \overline{u'v'} - \frac{r}{Re_b} \frac{\mathrm{d} \overline{u}}{\mathrm{d} r} \right).
\end{equation}
The relation among the skin-friction coefficients $C_{f,i}, C_{f,o}$ and the pressure gradient $\Pi_x$ is obtained by multiplying \eqref{eq:x-momentum} by $r$ and by integrating along $r$ between $R_i$ and $R_o$. As the Reynolds shear stresses $\overline{u'v'}$ vanish on the cylinders because of the no-slip condition, we obtain \citep{walker-etal-1957}
\begin{equation}
\label{eq:pi-x}
\Pi_x = \dfrac{R_o C_{f,o}-R_i C_{f,i}}{R_o^2 - R_i^2}.
\end{equation}
Henceforth, the terminology Reynolds stresses is adopted to denote $\overline{u'v'}$.
Substituting $\mathcal{D}^*=2 \pi L^* \mu^* \left(R_i^*  \mathrm{d} \overline{u}^*/\mathrm{d} r^*|_{r^*=R_i^*} - R_o^* \mathrm{d} \overline{u}^*/\mathrm{d} r^*|_{r^*=R_o^*}\right)$ (where $\mu^*$ is the dynamic viscosity of the fluid and $L^*$ is the length of the annulus) and $\mathcal{A}^*=2 \pi L^*\left(R_i^*+R_o^*\right)$ into the definition of the average $C_f$ leads to
\begin{equation}
\label{eq:cf-cfi-cfo}
C_f = \frac{R_i C_{f,i} - R_o C_{f,o}}{R_i+R_o}.
\end{equation}
By comparing \eqref{eq:pi-x} and \eqref{eq:cf-cfi-cfo} and using $R_o-R_i=1$, it follows that $C_f=-\Pi_x$.

\subsection{Laminar-flow quantities}

The laminar solution is obtained by integrating the $x$-momentum equation \eqref{eq:x-momentum} along $r$ with $\overline{u'v'}=0$ and by using the no-slip conditions, $\overline{u} = 0$ at $r=R_i, R_o$.
The solution is
\begin{equation}
\label{eq:laminar-profile}
\overline{u}_{lam}(r) = \dfrac{\Pi_{x,lam} Re_b}{4} \left[r^2 - R_i^2 + \frac{R_o^2 - R_i^2}{\ln\left(R_i/R_o\right)} \ln\left(\frac{r}{R_i}\right)\right],
\end{equation}
first derived by \citet[p.~586]{lamb-1932}. By substituting \eqref{eq:laminar-profile} into \eqref{eq:inner-cf-def} and \eqref{eq:outer-cf-def}, the laminar skin-friction coefficients are found:
\begin{equation}
\label{eq:cf-lam-pix}
C_{f,i,lam} = \Pi_{x,lam} \left[R_i + \frac{R_o^2 - R_i^2}{2R_i\ln\left(R_i/R_o\right)}\right], \quad C_{f,o,lam} = \Pi_{x,lam} \left[R_o + \frac{R_o^2 - R_i^2}{2R_o\ln\left(R_i/R_o\right)}\right].
\end{equation}
Substituting the laminar solution \eqref{eq:laminar-profile} into the bulk-velocity equation \eqref{eq:bulk-pressure}, we obtain
\begin{equation}
\label{eq:pix-lam}
\Pi_{x,lam} = - C_{f,lam} = \frac{8\ln\left(R_o/R_i\right)}{Re_b\left[R_o^2 - R_i^2 - (R_o^2 + R_i^2)\ln\left(R_o/R_i\right)\right]}.
\end{equation}
By combining \eqref{eq:cf-lam-pix} and \eqref{eq:pix-lam}, we find
\begin{equation}
\label{eq:cf-i-lam}
C_{f,i,lam} =  \frac{4\left[R_i^2 - R_o^2 + 2 R_i^2 \ln(R_o/R_i)\right]}{Re_b R_i \left[R_o^2 - R_i^2 - (R_o^2+R_i^2) \ln(R_o/R_i)\right]},
\end{equation}
\begin{equation}
\label{eq:cf-o-lam}
C_{f,o,lam} =  \frac{4\left[R_i^2 - R_o^2 + 2 R_o^2 \ln(R_o/R_i)\right]}{Re_b R_o \left[R_o^2 - R_i^2 - (R_o^2+R_i^2) \ln(R_o/R_i)\right]}.
\end{equation}
In terms of the radius ratio $\xi=R_o/R_i$ (thus using $R_i=1/(\xi-1)$ and $R_o=\xi/(\xi-1)$), the laminar pressure gradient and skin-friction coefficients are
\begin{equation}
\label{eq:pix-x-laminar-ratio}
\Pi_{x,lam} = - C_{f,lam} = \frac{8 (\xi-1)^2 \ln \xi}{Re_b\left[\xi^2 - 1 - (\xi^2 + 1)\ln \xi\right]},
\end{equation}
\begin{equation}
\label{eq:cf-i-lam-ratio}
C_{f,i,lam} = \frac{4(\xi-1)\left(1 - \xi^2 + 2 \ln \xi \right)}{Re_b \left[\xi^2 - 1 - \left(\xi^2+1\right) \ln \xi \right]},
\end{equation}
\begin{equation}
\label{eq:cf-o-lam-ratio}
C_{f,o,lam} = \frac{4(\xi-1)\left(1 - \xi^2 + 2 \xi^2 \ln \xi \right)}{Re_b \xi \left[\xi^2 - 1 - \left(\xi^2+1\right) \ln \xi \right]}.
\end{equation}
\subsection{Identities for the pressure gradient and the skin-friction coefficients}
\label{sec:identities}

The $x$-momentum equation \eqref{eq:x-momentum} is multiplied by $r$, integrated from $R_i$ to $r$ and each term is divided by $r$ to single out $\mathrm{d} \overline{u}/\mathrm{d} r$:
\begin{equation}
\label{eq:x-mom-int-3}
    - \frac{\Pi_x}{2}\left( r - \frac{R_i^2}{r} \right)
    =
    \overline{u'v'} - \frac{1}{Re_b} \frac{\mathrm{d} \overline{u}}{\mathrm{d} r}
    + \frac{R_i C_{f,i}}{2 r}.
\end{equation}
Equation \eqref{eq:x-mom-int-3} is integrated from $R_i$ to $r$ to obtain $\overline{u}$:
\begin{equation}
\label{eq:x-mom-int-6}
    - \frac{\Pi_x}{4} \left(r^2 - R_i^2\right) +
    \frac{\Pi_x R_i^2}{2} \ln\left( \frac{r}{R_i} \right)
    =
    \underbrace{\int_{R_i}^r \overline{u'v'} \mathrm{d}\tilde{r}}_{A}
    \underbrace{- \frac{\overline{u}}{Re_b}}_{B}
    + \frac{R_i C_{f,i}}{2} \ln\left( \frac{r}{R_i} \right).
\end{equation}
Equation \eqref{eq:x-mom-int-6} is multiplied by $r$ and integrated between $R_i$ and $R_o$. The integrated term A in \eqref{eq:x-mom-int-6} is simplified by using integration by parts. The bulk-velocity equation \eqref{eq:bulk-pressure} is used in the integration of term B in \eqref{eq:x-mom-int-6}. Equation \eqref{eq:x-mom-int-6} becomes
\begin{equation}
\label{eq:cfh-1}
\begin{split}
&
- \Pi_x \left( R_o^2 - R_i^2 \right)^2 + 2 R_i \left(\Pi_x R_i - C_{f,i}\right) \left[ R_i^2 - R_o^2 + 2 R_o^2\ln\left( \frac{R_o}{R_i} \right) \right]
=
\\ &
8 \int_{R_i}^{R_o} \overline{u'v'} \left( R_o^2 - r^2 \right) \mathrm{d} r
+ 8 \frac{R_i^2 - R_o^2}{Re_b}.
\end{split}
\end{equation}
By integrating \eqref{eq:x-mom-int-3} between $R_i$ and $R_o$ and by using the no-slip conditions on the cylinders, one finds
\begin{equation}
\label{eq:cfh-2}
-\dfrac{\Pi_x}{4}
\left[R_o^2 - R_i^2 - 2 R_i^2 \ln \left(\dfrac{R_o}{R_i}\right) \right]
=
\int_{R_i}^{R_o} \overline{u'v'} \mathrm{d}r+\dfrac{R_i C_{f,i}}{2} \ln \left(\dfrac{R_o}{R_i}\right).
\end{equation}
Combining \eqref{eq:cf-cfi-cfo}, \eqref{eq:cfh-1} and \eqref{eq:cfh-2} yields
\begin{equation}
\label{eq:cf-turbulent}
C_f = -\Pi_x = - \Pi_{x,lam} +
\frac{
4\left( R_o^2 - R_i^2 \right) \int_{R_i}^{R_o} \overline{u'v'} \mathrm{d} r
- 8 \ln \left(R_o/R_i\right) \int_{R_i}^{R_o} \overline{u'v'} r^2 \mathrm{d} r
}{\left[R_o^2 - R_i^2 - (R_o^2+R_i^2) \ln \left(R_o/R_i\right)\right]\left(R_o^2 - R_i^2\right)},
\end{equation}
\begin{equation}
\label{eq:cfi-turbulent}
\begin{split}
C_{f,i} = C_{f,i,lam} +
\frac{
4 \left[ R_i^2 - R_o^2 + 2 R_i^2 \ln \left(R_o/R_i\right) \right] \int_{R_i}^{R_o} \overline{u'v'}r^2 \mathrm{d} r +
2 (R_o^2-R_i^2)^2 \int_{R_i}^{R_o} \overline{u'v'} \mathrm{d} r
}{
\left[R_o^2 - R_i^2 - (R_o^2+R_i^2) \ln \left(R_o/R_i\right) \right]\left(R_o^2 - R_i^2\right)R_i
},
\end{split}
\end{equation}
\begin{equation}
\label{eq:cfo-turbulent}
\begin{split}
C_{f,o} = C_{f,o,lam} +
\frac{
         4 \left[ R_i^2 - R_o^2 + 2 R_o^2 \ln \left(R_o/R_i\right) \right] \int_{R_i}^{R_o} \overline{u'v'} r^2 \mathrm{d} r
        -2(R_o^2-R_i^2)^2\int_{R_i}^{R_o} \overline{u'v'} \mathrm{d} r
}
{
    \left[R_o^2 - R_i^2 - (R_o^2+R_i^2) \ln \left(R_o/R_i\right) \right] (R_o^2 - R_i^2) R_o
}.
\end{split}
\end{equation}
In terms of the radius ratio $\xi$, identities \eqref{eq:cf-turbulent}-\eqref{eq:cfo-turbulent} become
\begin{equation}
\label{eq:cf-ratio}
C_f = -\Pi_x = - \Pi_{x,lam} +
\frac{
4\left( \xi^2 - 1 \right)\int_{1}^{\xi} \overline{u'v'} \mathrm{d}\tilde r
- 8 \ln \xi \int_{1}^{\xi} \overline{u'v'} {\tilde r}^2 \mathrm{d}\tilde r
}{(\xi + 1)\left[\xi^2 - 1 - (\xi^2+1) \ln \xi \right]},
\end{equation}
\begin{equation}
\label{eq:cfi-turbulent-ratio}
\begin{split}
C_{f,i} = C_{f,i,lam} +
\frac{
4 \left( 1 - \xi^2 + 2 \ln \xi \right) \int_{1}^{\xi} \overline{u'v'} \tilde r^2 \mathrm{d}\tilde r
+
2 (\xi^2-1)^2 \int_{1}^{\xi} \overline{u'v'} \mathrm{d}\tilde r
}{
\left(\xi^2 - 1\right)\left[\xi^2 - 1 - (\xi^2+1) \ln \xi \right]
},
\end{split}
\end{equation}
\begin{equation}
\label{eq:cfo-turbulent-ratio}
\begin{split}
C_{f,o} = C_{f,o,lam} +
\frac{
         4 \left( 1 - \xi^2 + 2 \xi^2 \ln \xi \right) \int_{1}^{\xi} \overline{u'v'} \tilde r ^2 \mathrm{d}\tilde r
         -2(\xi^2-1)^2 \int_{1}^{\xi} \overline{u'v'} \mathrm{d}\tilde r
}
{
    \xi (\xi^2 - 1) \left[\xi^2 - 1 - (\xi^2+1) \ln \xi \right]
},
\end{split}
\end{equation}
where $\tilde r = r/R_i$.
\begin{figure}
\centering
\includegraphics[width=13.5cm]{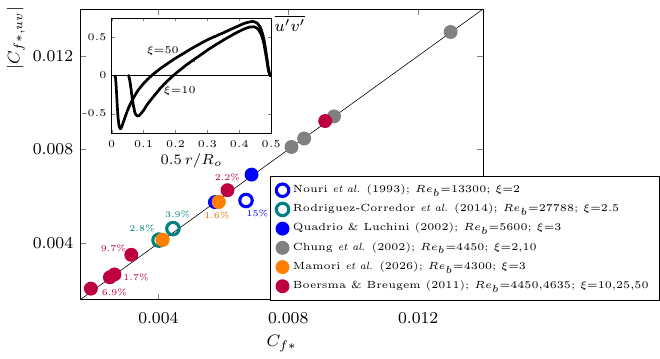}
\vspace{-4mm}
\caption{Absolute values of the skin-friction coefficients obtained via the theoretical formulas, $|C_{f*,uv}|$ (the subscript $*$ indicates any of \eqref{eq:cf-turbulent}, \eqref{eq:cfi-turbulent} or \eqref{eq:cfo-turbulent}), versus experimental (open symbols) and numerical (filled symbols) skin-friction coefficients reported in the literature, $C_{f*}$.
The coefficients $C_{f*}$ are shown without an absolute value because all cited studies report them using an unsigned (positive) convention. The colour-coded percent values quantify the differences between $|C_{f*,uv}|$ and $C_{f*}$ when the difference is larger than 1\%. The data of \citet{boersma-breugem-2011} are divided by three to avoid an excessively large range of values and preserve visual clarity. Parameter $\xi=R_o/R_i$ is the radius ratio. All data are for stationary cylinders, except those of \citet{mamori-etal-2026}, which are for oscillating cylinders. Inset: Reynolds stresses $\overline{u'v'}$ computed by \citet{boersma-breugem-2011} for $Re_b=9270$ ($\xi=$10) and $Re_b=8900$ ($\xi=$50).}
\label{fig:cf-vs-cf}
\end{figure}
The integral term $\int_{R_i}^{R_o} \overline{u'v'} \mathrm{d} r$ in \eqref{eq:cf-turbulent}-\eqref{eq:cfo-turbulent} is not zero because $\overline{u'v'}$ is not symmetric with respect to the midplane between the two cylinders, as shown by the numerical data of \citet{boersma-breugem-2011} in the inset of figure \ref{fig:cf-vs-cf} (refer also to figure 5 of \citet{quadrio-luchini-2002} and figure 20 of \citet{rodriguez-etal-2014}). The profiles in figure \ref{fig:cf-vs-cf} further reveal that, as the radius ratio $\xi$ increases and the annular flow approaches a pipe flow, the negative peak becomes more pronounced and the point where $\overline{u'v'}=0$ moves closer to the inner cylinder, thereby making the positive area under the curve larger.

Equations \eqref{eq:cf-turbulent}-\eqref{eq:cfo-turbulent} are also valid when either or both cylinders are in motion along the azimuthal direction because the azimuthal velocity component is absent from the averaged $x-$momentum equation \eqref{eq:x-momentum}.
The Reynolds-number dependence of the skin-friction coefficients expressed in identities \eqref{eq:cf-turbulent}-\eqref{eq:cfo-turbulent} is found explicitly in the laminar parts $C_{f,lam}$, $C_{f,i,lam}$, $C_{f,o,lam}$, given in \eqref{eq:pix-lam}-\eqref{eq:cf-o-lam}, and implicitly in the Reynolds-stress integrals in \eqref{eq:cf-turbulent}-\eqref{eq:cfo-turbulent}. Although the Reynolds stresses $\overline{u'v'}$ tend to agree more closely with the total stress in the core of the annular gap as the Reynolds number increases, at any Reynolds number the near-wall viscous effects exert an influence on $\overline{u'v'}$, and therefore on $C_{f}$, $C_{f,i}$, $C_{f,o}$ through the integrals \eqref{eq:cf-turbulent}-\eqref{eq:cfo-turbulent}. The accurate quantification of the near-wall behaviour of $\overline{u'v'}$ thus establishes the Reynolds-number dependence of $C_{f}$, $C_{f,i}$, $C_{f,o}$. The neglect of these viscous effects leads to constant $C_{f}$, $C_{f,i}$, $C_{f,o}$ at a large Reynolds number, which contradicts experimental and numerical evidence.

Figure \ref{fig:cf-vs-cf} reports the absolute values of the skin-friction coefficients computed using the theoretical formulas \eqref{eq:cf-turbulent}-\eqref{eq:cfo-turbulent} versus the corresponding coefficients reported in the literature. We selected datasets that include both $\overline{u'v'}$ and $C_f$. The Reynolds-stress data were digitised from the figures reported in the studies cited in figure \ref{fig:cf-vs-cf}. The trapezoidal rule was used for the computation of the integrals in identities \eqref{eq:cf-turbulent}-\eqref{eq:cfo-turbulent} and in the following \eqref{eq:rm-turbulent-2}, \eqref{eq:cf-w} and \eqref{eq:ubw-solution}.
The agreement is excellent for most of the experimental and numerical datasets. \citet{quadrio-luchini-2002}, \citet{chung-etal-2002} and \citet{boersma-breugem-2011} (for $\xi=50$) (stationary cylinders) show differences below 1\%. The data of \citet{mamori-etal-2026} present a maximum difference of 1.6\% (oscillating cylinders). The data of \citet{boersma-breugem-2011} for $\xi=10$ and $\xi=25$ show 9.7\% and 2.2\% differences, respectively. The experimental data of \citet{rodriguez-etal-2014} present a maximum difference of 3.9\% and those by \citet{nouri-etal-1993} show a disparity of 15\%, arguably due to the uncertainty of the measured Reynolds stresses in their figure 2(c).

The limit of vanishingly thin inner cylinder, i.e. $R_i^*/h^*\rightarrow 0$ ($\xi \rightarrow \infty$, or $R_o\rightarrow 1$ and $R_i \rightarrow 0$), leads to
\begin{align}
C_f
&
\sim -\frac{8\ln\left(R_o/R_i\right)}{Re_b\left[1 - \ln\left(R_o/R_i\right)\right]} +
\frac{
4 \int_{R_i}^{R_o} \overline{u'v'} \mathrm{d} r
- 8 \ln \left(R_o/R_i\right) \int_{R_i}^{R_o} \overline{u'v'} r^2 \mathrm{d} r
}{1 - \ln \left(R_o/R_i\right)}
\\ &
\sim \frac{8}{Re_b} + 8 \int_0^1 \overline{u'v'} r^2 \mathrm{d} r \sim -C_{f,o}.
\label{eq:pipe}
\end{align}
After rescaling, \eqref{eq:pipe} coincides with equation (19) in FIK for circular pipe flow despite the streamwise velocity $\overline{u}$ satisfying different conditions at the centreline in the pipe-flow case and in the annular-flow case as $R_i \rightarrow 0$. In this limit, although the wall-shear stress at the centreline grows unbounded, the surface area vanishes. It follows that the friction drag on the inner cylinder vanishes $\propto (\ln \xi)^{-1}$ as $\xi \rightarrow \infty$. The total drag is thus given by the friction on the outer cylinder only.

The limit of vanishingly thin annular gap, i.e. $h^*/R_o^* \rightarrow 0$ ($\xi \rightarrow 1$, or $R_i, R_o \rightarrow \infty$ with $R_o-R_i=1$), is studied to verify that \eqref{eq:cf-turbulent}-\eqref{eq:cfo-turbulent} simplify to the identity for fully developed planar channel flow discovered by FIK. This limit is obtained using the Taylor series $\ln(\epsilon+1) = \epsilon - \epsilon^2/2 + \epsilon^3/3...,$ for $|\epsilon|\ll1,$ where $R_o/R_i=\epsilon+1$. Substituting the resulting expansion $\ln\left(R_o/R_i\right) \sim (6 R_i^2 - 3 R_i + 2)/(6 R_i^3)$ and the change of variable $y=r-R_i=\mathcal{O}(1)$ into \eqref{eq:pix-lam}-\eqref{eq:cf-o-lam} and \eqref{eq:cf-turbulent}-\eqref{eq:cfo-turbulent}, and using $\int_0^1 \overline{u'v'} \mathrm{d}y=0$ arising from the symmetry of the channel flow with respect to the channel midplane, it follows that $C_{f_i}\sim-C_{f_o}\sim C_f$, where
\begin{equation}
\label{eq:channel}
C_f \sim \frac{12}{Re_b} - 12\int_0^1 \left( \dfrac{1}{2} - y \right) \overline{u'v'} \mathrm{d}y.
\end{equation}
After rescaling, \eqref{eq:channel} coincides with equation (18) in FIK for planar channel flow.

An identity for the averaged skin-friction coefficient for annular flows has recently been derived by \citet{mamori-etal-2026}. The main differences with respect to our \eqref{eq:cf-turbulent} are that, in the identity of \citet{mamori-etal-2026}, (i) the laminar skin-friction coefficient is not isolated; (ii) the integral terms involve the radial location $r_m$ of the maximum streamwise velocity, which appears in the integrands, in the integration limits and in other {multiplicative } factors; (iii) the wall-normal integration is conducted separately on different integrands from the inner cylinder to $r_m$ and from $r_m$ to the outer cylinder; (iv) the identity is implicit with respect to the skin-friction coefficients because $r_m$ is expressed as a function of $C_{f,i}$ and $C_{f,o}$; (v) a term $C_{f,bulk}$ appears, which involves integrals of the streamwise mean velocity and finds no equivalent in either \eqref{eq:cf-turbulent} or the classical FIK identities.

\subsection{Identities for the radial location of the maximum mean velocity}

\begin{figure}
\centering
\includegraphics[width=6.7cm]{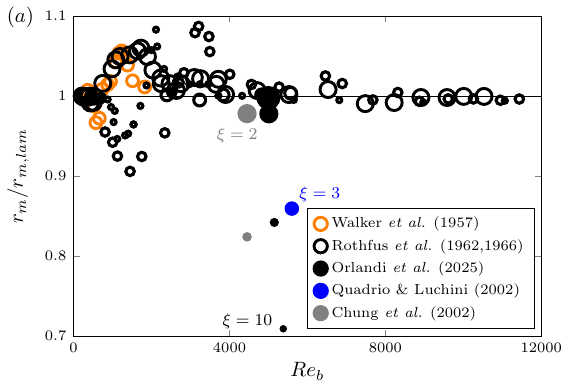}
\includegraphics[width=6.7cm]{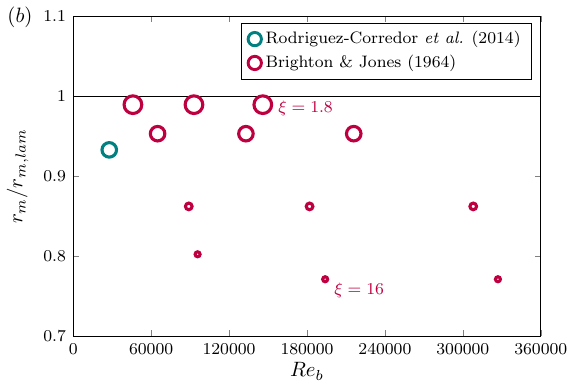}
\vspace{-6mm}
\caption{Ratio between the radial locations $r_m$ of the maximum $\overline{u}$ and the corresponding location $r_{m,lam}$ in the laminar case. (a) $Re_b<12000$; (b) $Re_b>45000$. Filled circles: numerical data; open circles: experimental data. The symbol size is inversely proportional to the radius ratio $\xi=R_o/R_i$ (colour-coded,  representative $\xi$ values are given). In the legend of panel ($a$), Rothfus {\em et al.} (1962, 1966) indicates \citet{croop-rothfus-1962} and \citet{rothfus-etal-1966}.}
\label{fig:rm-lam-turb}
\end{figure}

It is also instructive to study how the radial location $r_m$ of the maximum mean streamwise velocity varies for different flow conditions. Figure \ref{fig:rm-lam-turb} presents numerical and experimental data for the ratio $r_m/r_{m,lam}$, where $r_{m,lam}$ is the radial location of the laminar streamwise velocity, i.e. $r_{m,lam}=\left[\left(R_o^2 - R_i^2\right)/[2 \ln \left(R_o/R_i\right)]\right]^{1/2}$. In the transitional regime, the results by Rothfus and co-workers \citep{walker-etal-1957,croop-rothfus-1962} demonstrate that the ratio can be smaller or larger than unity. In the turbulent regime, except for the data of Rothfus and co-workers (e.g. \citeauthor{rothfus-etal-1966} \citeyear{rothfus-etal-1966}), consensus exists on $r_m/r_{m,lam}$ being smaller than unity and decreasing with $R_i/R_o$, i.e. $r_m$ becomes smaller than $r_{m,lam}$ as both quantities approach the centreline of the annular system and the ratio $R_i/R_o$ decreases. Figure \ref{fig:rm-lam-turb} further indicates that the ratio becomes independent of the Reynolds number as the Reynolds number increases, a result that is particularly evident from the experimental data of \citet{brighton-jones-1964}, conducted at the highest Reynolds numbers to date.

As a first step to find a relation between the radial location $r_m$ of the maximum mean streamwise velocity and the integrated Reynolds stresses, it is useful to discuss first whether the Reynolds stresses vanish at $r_m$, a controversial point in the study of annular flows. The experimental results of \citet{lawn-elliott-1972}, \citet{rehme-1974}, \citet{rodriguez-etal-2014} and \citet{ghaemi-etal-2015} showed that the radial locations where $\overline{u'v'}=0$ and $\mathrm{d}\overline{u}/\mathrm{d}r=0$ did not coincide, implying that, in a small range of radial locations, negative production of turbulent kinetic energy occurred. \citet{ghaemi-etal-2015} discussed a few physical interpretations for this occurrence, but also hypothesised that the flow might not have reached fully developed conditions. The direct numerical simulations of \citet{chung-etal-2002} reported a mismatch between the two locations, whereas, in the numerical studies of \citet{quadrio-luchini-2002}, \citet{boersma-breugem-2011} and \citet{bagheri-wang-2020}, the two values coincided. \citet{boersma-breugem-2011} claimed that experimental errors might have led to the conclusions of the experimental studies and attributed the differences between the two locations found by \citet{chung-etal-2002} to insufficient spatial resolution. Highly accurate direct numerical simulations were therefore conducted by \citet{orlandi-etal-2025} for a wide range of radius ratios to resolve this issue. In all those cases, the locations where $\overline{u'v'}=0$ and $\mathrm{d}\overline{u}/\mathrm{d}r=0$ coincided exactly.

The location $r_m$ is thus found by setting $\overline{u'v'}=0$ and $\mathrm{d}\overline{u}/\mathrm{d}r=0$ in \eqref{eq:x-mom-int-3} and by using \eqref{eq:pi-x} \citep{bagheri-wang-2020},
\begin{equation}
\label{eq:rm-turbulent}
r_m = \left[R_i^2 - \dfrac{R_iC_{f,i}}{\Pi_x}\right]^{1/2} = \left[R_i R_o \dfrac{R_i C_{f,o}-R_o C_{f,i}}{R_o C_{f,o}-R_i C_{f,i}}\right]^{1/2}.
\end{equation}
Substituting \eqref{eq:cf-turbulent} and \eqref{eq:cfi-turbulent} into \eqref{eq:rm-turbulent} gives $r_m$ as a function of the integrated Reynolds stresses,
\begin{equation}
\label{eq:rm-turbulent-2}
r_m=
\left[
\frac{
(R_o^2-R_i^2)
\left[
2\left(R_o^2-R_i^2\right)
+2Re_b \int_{R_i}^{R_o}\overline{u'v'} r^2 \mathrm{d}r
-Re_b (R_i^2 + R_o^2)
 \int_{R_i}^{R_o}\overline{u'v'} \mathrm{d}r
\right]
}{
4\ln\left(R_o/R_i\right)
\left(
R_o^2-R_i^2
+Re_b \int_{R_i}^{R_o}\overline{u'v'}r^2 \mathrm{d}r
\right)-
2Re_b(R_o^2-R_i^2)
\int_{R_i}^{R_o}\overline{u'v'} \mathrm{d}r
}\right]^{1/2},
\end{equation}
which reduces to the laminar $r_{m,lam}$ when $\overline{u'v'}$=0.
\begin{figure}
\centering
\includegraphics[width=10.5cm]{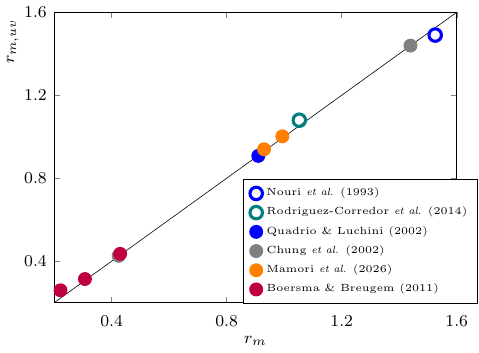}
\vspace{-4mm}
\caption{Radial locations $r_{m,uv}$ of the maximum $\overline{u}$ computed by the theoretical identity \eqref{eq:rm-turbulent-2} versus experimental (open symbols) and numerical (filled symbols) radial locations $r_m$.}
\label{fig:rm-vs-rm}
\end{figure}
In terms of the radius ratio $\xi$, identity \eqref{eq:rm-turbulent-2} becomes
\begin{equation}
\label{eq:rm-turbulent-2-ratio}
r_m^2=
\mathcal{D}_m
\left[
2(\xi^2-1)^2+2Re_b (\xi+1) \int_{1}^{\xi}\overline{u'v'} \tilde r^2 \mathrm{d}\tilde r
-Re_b(\xi^2+1)(\xi+1) \int_{1}^{\xi}\overline{u'v'}d \tilde r
\right],
\end{equation}
where
\begin{equation}
\begin{aligned}
\mathcal{D}_m(\xi)=
&
\left[4\ln\xi \left[(\xi+1)(\xi-1)^3+Re_b(\xi-1) \int_{1}^{\xi}\overline{u'v'}\tilde r^2 \mathrm{d}\tilde r\right]
\right.
\\ &
\left.
-2Re_b (\xi+1) (\xi-1)^2\int_{1}^{\xi}\overline{u'v'}\mathrm{d}\tilde r
\right]^{-1}.
\end{aligned}
\end{equation}
Figure \ref{fig:rm-vs-rm} compares the $r_m$ data computed via \eqref{eq:rm-turbulent-2} with data in the literature. The percentage deviations from our computations of $r_m$ are smaller than or comparable with those reported in figure \ref{fig:cf-vs-cf} for the skin-friction coefficients. By substituting $r_m$ into \eqref{eq:x-mom-int-6}, the maximum mean velocity $u_m$ is determined as a function of $R_i, R_o, Re_b$ and $\overline{u'v'}$.

The result that $r_m/r_{m,lam}$ is smaller than unity in the turbulent regime, shown in figure \ref{fig:rm-lam-turb}, is confirmed by the asymptotic form of \eqref{eq:rm-turbulent-2} in the limit $R_i\rightarrow 0$ with $R_o=O(1)$,
\begin{equation}
\label{eq:rm-turbulent-3}
\begin{aligned}
r_m^2
\sim&
\underbrace{\dfrac{R_o^2 - R_i^2}{2\ln\!\left(R_o/R_i\right)}}_{r_{m,lam}^2}
-\underbrace{\dfrac{Re_b (R_o^4 - R_i^4) \int_{R_i}^{R_o} \overline{u'v'} \,\mathrm{d}r}{4 \left(R_o^2 - R_i^2 + Re_b \int_{R_i}^{R_o} \overline{u'v'}\, r^2 \,\mathrm{d}r\right)\ln\left(R_o/R_i\right)}}_{\mathcal{R}_m}.
\end{aligned}
\end{equation}
The quantity $\mathcal{R}_m$ in \eqref{eq:rm-turbulent-3} is positive because the integrals involving the Reynolds stresses are both positive when the flow is turbulent. This result is evident from available $\overline{u'v'}$ profiles at large radius ratio $\xi$, such as those shown in the inset of figure \ref{fig:cf-vs-cf}. The Reynolds stresses $\overline{u'v'}$ are positive for a wide range of radial positions in the outer portion of the annular gap and the positive $\overline{u'v'}$ peak is more intense than the negative peak \citep{boersma-breugem-2011,chung-etal-2002}. Therefore, the second term in \eqref{eq:rm-turbulent-3} renders $r_m/r_{m,lam}<1$ because of its negative sign.

\section{Results for cylinder-driven annular flow}

\subsection{Relations between the skin-friction coefficients}

The averaged $x$-momentum equation for the cylinder-driven annular flow is:
\begin{equation}
\label{eq:x-momentum-w}
    \frac{\mathrm{d}}{\mathrm{d} r}\left(r \overline{u'v'} - \frac{r}{Re_c} \frac{\mathrm{d} \overline{u}}{\mathrm{d} r} \right)=0.
\end{equation}
The relation among the skin-friction coefficients $C_{f,i,c}, C_{f,o,c}$ is obtained by integrating \eqref{eq:x-momentum-w} along $r$. As the Reynolds stresses $\overline{u'v'}$ vanish on the cylinders because of the no-slip condition, one finds $R_o C_{f,o,c}=R_i C_{f,i,c}$.

\subsection{Laminar-flow quantities}
The laminar solution for the cylinder-driven flow is obtained by integrating \eqref{eq:x-momentum-w} along $r$ with $\overline{u'v'}=0$ and by using the boundary conditions, $\overline{u}(R_i)=U_c$ and $\overline{u}(R_o)=0$ (where $U_c=1$, though left unassigned for clarity). The solution is $\overline{u}_{lam}(r)=U_c \ln\left(r/R_o\right)/\ln(R_i/R_o)$. The laminar skin-friction coefficients are:
\begin{equation}
\label{eq:cf-w-lam}
C_{f,i,c,lam}=\dfrac{-2 U_c}{R_i Re_c \ln(R_o/R_i)}, \quad C_{f,o,c,lam}=\dfrac{-2 U_c}{R_o Re_c \ln(R_o/R_i)}.
\end{equation}

\subsection{Identities for the skin-friction coefficients}

The $x$-momentum equation \eqref{eq:x-momentum-w} is integrated from $R_i$ to $r$ to single out $\mathrm{d} \overline{u}/\mathrm{d} r$:
\begin{equation}
\label{eq:x-mom-w-2}
\overline{u'v'} - \frac{1}{Re_c} \frac{\mathrm{d} \overline{u}}{\mathrm{d} r}
+ \dfrac{R_i C_{f,i,c}}{2 r}=0.
\end{equation}
By integrating \eqref{eq:x-mom-w-2} between $R_i$ and $R_o$ and by using \eqref{eq:cf-w-lam}, we obtain
\begin{equation}
\label{eq:cf-w}
C_{f,i,c}=C_{f,i,c,lam}-\dfrac{2 \int_{R_i}^{R_o} \overline{u'v'} \mathrm{d}r}{R_i \ln\left(R_o/R_i\right)}, \quad
C_{f,o,c}=C_{f,o,c,lam}-\dfrac{2 \int_{R_i}^{R_o} \overline{u'v'} \mathrm{d}r}{R_o \ln\left(R_o/R_i\right)}.
\end{equation}
The average skin-friction coefficient $C_{f,c}$ is found by substituting identities \eqref{eq:cf-w} into \eqref{eq:cf-cfi-cfo}, valid for the cylinder-driven case as well.

\begin{figure}
\centering
\includegraphics[width=12cm]{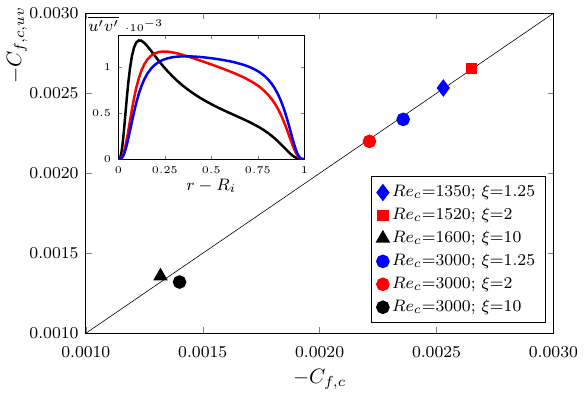}
\vspace{-4mm}
\caption{Skin-friction coefficients computed via the theoretical formulas \eqref{eq:cf-w} and \eqref{eq:cf-cfi-cfo}, $-C_{f,c,uv}$, versus the numerical skin-friction coefficients computed by \citet{kunii-etal-2019}, $-C_{f,c}$. Inset: Reynolds stresses $\overline{u'v'}$ computed by \citet{kunii-etal-2019} for $Re_c=3000$ for $\xi=1.25$ (blue), $\xi=2$ (red) and  $\xi=10$ (black).}
\label{fig:cf-vs-cf-kunii}
\end{figure}

It was numerically verified by \citet{kunii-etal-2019} that $\overline{u'v'}>0$ $\forall r$ (and, analogously, in plane Couette flows \citep{kawata-alfredsson-2019}), as shown in the inset of figure \ref{fig:cf-vs-cf-kunii}. It follows that the Reynolds-stress integral terms in \eqref{eq:cf-w} quantify the increase of the skin-friction coefficients in absolute value with respect to the laminar coefficients, as demonstrated by figure 12(a) of \citet{kunii-etal-2019} (where their Reynolds number is $Re=Re_c/4$). Figure \ref{fig:cf-vs-cf-kunii} also compares the average skin-friction coefficients computed using the theoretical formulas \eqref{eq:cf-w} and \eqref{eq:cf-cfi-cfo} with the average coefficients computed by \citet{kunii-etal-2019}. The agreement is excellent for all cases, except for the coefficient for $Re_c=3000$ and $\xi=10$ (black circle), showing a discrepancy of about 5\%.

\subsection{Identities for the bulk velocity}

\begin{figure}
\centering
\includegraphics[width=6.7cm]{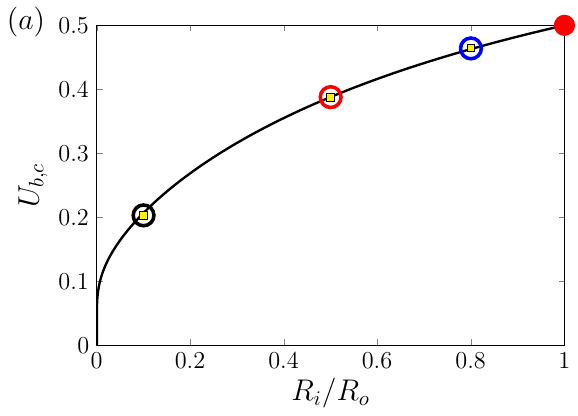}
\includegraphics[width=6.7cm]{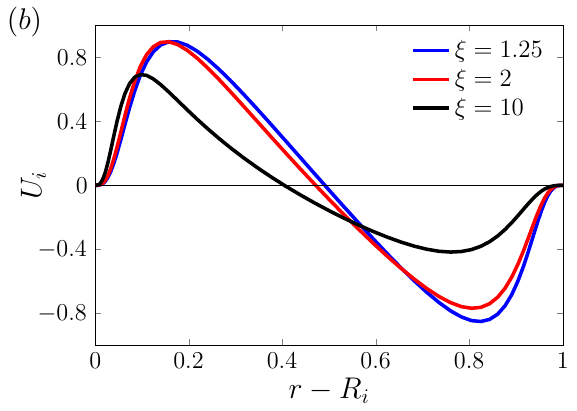}
\vspace{-6mm}
\caption{(a) Bulk velocity for cylinder-driven flow versus ratio $R_i/R_o$. The yellow squares denote data computed by integrating the $\overline{u}$ profiles by \citet{kunii-etal-2019}; the filled red circle denotes the bulk velocity for Couette flow \citep{kawata-alfredsson-2019}; the open circles denote the data computed via the theoretical formula \eqref{eq:ubw-solution} for $Re_c=3000$, coloured to match the data in panel (b); the solid line represents the laminar bulk velocity $U_{b,c,lam}$, given in \eqref{eq:ubw-solution}. (b) Integrand $U_i$, given in \eqref{eq:ubw-solution}, for $Re_c=3000$. Parameter $\xi=R_o/R_i$ is the radius ratio.}
\label{fig:kunii-bulk-velocity}
\end{figure}

By using \eqref{eq:bulk-cylinder} and \eqref{eq:cf-w-lam}-\eqref{eq:cf-w}, the bulk velocity is expressed as a function of the integrated Reynolds stresses,
\begin{equation}
\label{eq:ubw-solution}
U_{b,c}
=
\underbrace{
\dfrac{U_c \left[R_o^2 - R_i^2 - 2R_i^{2}\ln(R_o/R_i)\right]}{2\left(R_o^{2}-R_i^{2}\right)\ln(R_o/R_i)}
}_{U_{b,c,lam}}
+
\int_{R_i}^{R_o}
\underbrace{
\dfrac{Re_c \left[R_o^2-R_i^2 - 2 r^2 \ln(R_o/R_i) \right]}{2 \left(R_o^2-R_i^2\right) \ln(R_o/R_i)}
\overline{u'v'}
}_{U_i}
\mathrm{d}r.
\end{equation}
In terms of the radius ratio $\xi$, identity \eqref{eq:ubw-solution} becomes
\begin{equation}
\label{eq:ubw-solution-ratio}
U_{b,c}
=
\dfrac{U_c \left(\xi^2 - 1 - 2\ln\xi\right)}{2\left(\xi^2-1\right)\ln\xi}
+
\int_{1}^{\xi}
\dfrac{Re_c
\left(\xi^2 - 1 - 2 \tilde r ^2 \ln\xi \right)
}{2 \left(\xi-1\right) \left(\xi^2-1\right) \ln \xi}
\overline{u'v'}
\mathrm{d} \tilde r.
\end{equation}
The integral term in \eqref{eq:ubw-solution} quantifies the decrease of the bulk velocity with respect to the laminar $U_{b,c,lam}$, i.e. the Reynolds stresses have the effect of reducing the averaged streamwise flow rate produced by the motion of the inner cylinder with respect to the laminar value. This result is analogous to that observed in channel flow at constant pressure gradient when the flow transitions from the laminar to the turbulent regime \citep{marusic-joseph-mahesh-2007}.
Figure \ref{fig:kunii-bulk-velocity}(a), however, illustrates that the $U_{b,c,lam}$ values agree very well with the $U_{b,c}$ data in the range 1300$\leq$$Re_c$$\leq$3000, computed by integrating the streamwise velocity profiles of figure 12(a) in \citet{kunii-etal-2019} (yellow squares). The largest discrepancy between the bulk-velocity values occurs for $R_i/R_o=0.1$: $U_{b,c}$ is 1.7\% smaller than $U_{b,c,lam}$. This finding reveals that the decrease of mass flow rate due to the Reynolds stresses is negligible at these Reynolds numbers. The $U_{b,c}$ data computed by integrating $\overline{u}$ show excellent agreement with the $U_{b,c}$ data computed by the theoretical formula \eqref{eq:ubw-solution} ($Re_c$$=$3000, open coloured circles in figure \ref{fig:kunii-bulk-velocity}a). The differences are always smaller than 0.2\%, providing further evidence of the high-quality data computed by \citet{kunii-etal-2019}.
The reason for the negligible differences between $U_{b,c,lam}$ and $U_{b,c}$ at these Reynolds numbers lies in the profiles of the integrand $U_i$ in \eqref{eq:ubw-solution}, shown in figure \ref{fig:kunii-bulk-velocity}(b). These profiles are either nearly antisymmetric about the midpoint of the annular gap (for $\xi=1.25$ and $2$) or, if not, nevertheless exhibit nearly equal and opposite areas bounded by the curve at either side of the radial location where $U_i=0$ (for $\xi=10$).
Since $U_i$ is proportional to $Re_c$, an interesting line of future research is to investigate whether the integration of $U_i$ at larger $Re_c$ leads to a quantifiable difference between $U_{b,c}$ and $U_{b,c,lam}$.

The limits for a vanishingly thin annular gap are found following the procedure detailed in \S\ref{sec:identities},
\begin{equation}
\label{eq:cfw-limit-small-h}
C_{f,i,c} \sim C_{f,o,c} \sim -\dfrac{2 U_c}{Re_c}-2 \int_0^1 \overline{u'v'} \mathrm{d}y,
\end{equation}
\begin{equation}
\label{eq:ubw-plane-couette}
U_{b,c} \sim \dfrac{U_c}{2} + Re_c \int_0^1 \left( \dfrac{1}{2} - y\right) \overline{u'v'} \mathrm{d}y = \dfrac{U_c}{2}.
\end{equation}
After rescaling, \eqref{eq:cfw-limit-small-h} coincides with the identity discovered by \citet{kawata-alfredsson-2019} for planar Couette flow. The integral term in \eqref{eq:ubw-plane-couette} is null because, with respect to $y=1/2$, the integrand is odd as $\overline{u'v'}$ is even, i.e. the averaged flow rate is unaffected by the Reynolds stresses in the planar Couette flow case.

\section{Concluding remarks}
Novel integral identities for the skin-friction coefficients of annular flows, expressed in terms of the Reynolds stresses integrated along the radial direction, have been derived and used to analyse and assess the quality of experimental and numerical data available in the literature. Identities for the radial location of the maximum streamwise velocity in the pressure-driven case and for the bulk velocity in the cylinder-driven case have also been found for the first time. They have been instrumental in elucidating previously unexplained results on these quantities. The identities for the skin-friction coefficients could be useful for computing the friction drag once accurate models for the Reynolds stresses are available. At high Reynolds numbers, they could be especially useful when measurements of the streamwise pressure gradient are not available because the accuracy of the integrated Reynolds stresses is progressively less impacted by the error-carrying near-wall measurements as the Reynolds number increases. Other interesting areas of impact of these integral relations are the theory of bounds of the skin-friction coefficients and the flow control for a reduction of the streamwise pressure gradient. Thanks to these identities, the focus of these analyses rests solely on the Reynolds stresses in the bulk of the flow. Future lines of research could focus on extensions to heat-transfer cases and the inclusion of roughness effects.

\backsection[Acknowledgements]{The author would like to thank Professor T. Tsukahara (Tokyo University of Science) and Professor Y. Duguet (LISN-CNRS, Universit\'e Paris Saclay) for the interesting discussions about their paper \citep{kunii-etal-2019} and for sharing their numerical data of the Reynolds stresses and the skin-friction coefficients. To generate figure \ref{fig:cf-vs-cf-kunii}, their Reynolds-stress data were used in the computations of identities \eqref{eq:cf-w} and \eqref{eq:cf-cfi-cfo}, and their skin-friction-coefficient data were used as the comparison data. Their Reynolds-stress data were also used in the computation of identity \eqref{eq:ubw-solution} to generate figure \ref{fig:kunii-bulk-velocity}. The author has used the AI tool Claude for checking spelling and the references to equations, tables, figures and cited articles. No AI tools have been used in the extraction of data, mathematics and computing for this research.}
\vspace{-2mm}
\backsection[Funding]{The present study was not funded by any grant. However, background research carried out while preparing the proposal for grant UKRI3728 (subsequently funded by UKRI CRCRM, titled ``Radical novel design of a haemodialyser based on fluid mixing'') motivated the direction of this work.}
\vspace{-2mm}
\backsection[Declaration of interests]{The author reports no conflict of interest.}
\vspace{-5mm}
\bibliographystyle{jfm}
\bibliography{pr}

\end{document}